\documentstyle[aps,preprint,pre]{revtex}

\begin{document}
\draft

\title{\bf 
Layer by layer epitaxy in limited mobility nonequilibrium models
of surface growth
}

\author{
P. Punyindu Chatraphorn$^{(1,2)}$ and S. Das Sarma$^{(1)}$
}
\address{
$^{(1)}$Condensed Matter Theory Center,
Department of Physics, University of Maryland, College Park,
MD 20742-4111, USA \\
$^{(2)}$Department of Physics, Faculty of Science,
Chulalongkorn University, Bangkok 10330, Thailand}

\date{\today}
\maketitle  

\begin{abstract}
We study, using noise reduction techniques, layer by layer epitaxial
growth in limited mobility solid-on-solid nonequilibrium surface growth
models, which have been introduced in the context of kinetic surface
roughening in ideal molecular beam epitaxy.
Multiple hit noise reduction and long surface diffusion length lead
to qualitatively similar layer by layer epitaxy in 1+1 and 2+1 
dimensional limited mobility growth simulations. 
We discuss the dynamic scaling characteristics connecting the
transient layer by layer growth regime with the asymptotic 
kinetically rough growth regime.
\end{abstract}

\pacs{PACS: 05.70.Ln, 68.55.-a, 64.60.My, 81.15.Aa, 68.55.Jk, 68.35.Bs}

\vskip 1pc
\vfill\eject

\section{Introduction}

Thin film growth, under solid-on-solid epitaxial conditions, from the
vacuum vapor deposition of an atomic or a molecular beam (the so-called
molecular beam epitaxy or MBE) is an important technological process
used extensively to produce high quality thin films with smooth and
flat surfaces and interfaces. It is also a growth process of considerable
fundamental significance \cite{1,2,3,4,5,6} in the statistical 
mechanics of nonequilibrium phenomena because MBE (at least in its ideal
form \cite{7}, with no evaporation and vacancy or overhang formation
at the growth front) in principle represents \cite{7,8} a novel 
universality class of nonlinear surface growth outside the generic
Kardar-Parisi-Zhang universality \cite{9}.
A great deal of attention has therefore focused over the last ten years
on the statistical properties of kinetically rough (and, in principle,
generically scale invariant) surface growth in low temperature 
(room temperature or below) MBE, following the suggestions
\cite{7,8,10,11,12} of the possible importance of ideal MBE in defining 
novel growth universality classes in kinetic surface roughening
\cite{1,2,3,4,5,6}.
We note that the conserved surface current nature \cite{7,8}
of ideal MBE growth (i.e. solid-on-solid growth with no evaporation
and vacancy or overhang formation) rules out a KPZ description of
its growth dynamics.

It is interesting, perhaps even ironic, that MBE growth has played such 
a central role \cite{1,2,3,4,5,6} in kinetic surface roughening 
phenomena because the primary materials science  impetus for MBE
growth is obviously to avoid kinetic roughening as completely as
possible in order to produce smooth and flat thin films of high
surface quality with minimal amount of surface roughness. From the
materials science perspective of producing high quality smooth
(i.e. manifestly non-rough) thin films, therefore, MBE is typically
carried out at elevated temperatures ($\sim$ 500-1000 K), where
fast surface diffusion enables one to produce smooth thin films with
very little surface roughness.
Smooth MBE growth, as opposed to kinetically rough (low temperature) 
growth, is characterized 
by layer-by-layer growth oscillations \cite{13,14}, where each layer
of the growing thin film (on a singular high symmetry substrate)
essentially fills up completely before the next layer deposition
begins (on the other hand, kinetically rough growth is, by definition,
multilayer as many layers are partially filled at the growth front
producing increasing surface roughness with {\it no} layer-by-layer
oscillations),
consequently the surface morphology and associated properties 
oscillate as growth proceeds.
In the ideal layer-by-layer growth mode, therefore, the interface
width ($W$, the root mean square fluctuation in the interface height)
of the growing film oscillates (nominally between 0 and 1, as
measured in lattice units, indicating an empty or a filled layer)
whereas in the kinetically rough growth mode $W$ increases monotonically
as a power law in the average film thickness ($\langle h \rangle$).
Layer-by-layer epitaxy is, however, an initial transient growth 
regime which eventually crosses over to the asymptotic kinetically
rough growth regime as the shot noise intrinsic in the incident
deposition beam fluctuations always wins out to damp the layer-by-layer
oscillations, and at long enough time scales (and for large enough
lateral system sizes) statistically scale invariant kinetically rough
growth would always emerge.
(In fact, the noise associated with the stochastic diffusion process
also contributes to kinetic roughening, but the shot noise 
associated with the incident beam fluctuations is the most
{\it relevant} roughening mechanism.)
This is also the experimental observation :
layer-by-layer growth oscillations, as studied for example 
through RHEED intensity oscillations monitoring the dynamical surface
evolution \cite{13,14}, eventually always damp out as the stochastic
depositopn shot noise associated with incident particle beam 
fluctuations leads to kinetically rough multilayer growth after some 
characteristic time $t_c$.
The damping time $t_c$, beyond which layer by layer growth dies out,
depends on the growth temperature (which controls the surface 
diffusion rate), and is in general larger for higher temperatures
because longer diffusion lengths at higher temperatures enhance
layer by layer growth.
(There is actually a complication, arising from the unaviodable 
vicinality in the starting substrate which can never really be
precisely a high symmetry singular plane in real growth where 
layer by layer oscillations tend to disappear at both high and low
growth temperatures --- the low temperature behavior is from the
multilayer kinetically rough growth as discussed above, but the
high temperature disappearance arises from the so-called step
flow growth mode which is caused by the very fast surface diffusion
of deposited atoms at high temperatures leading to their moving
directly to step edges, which must be present in any real substrate
due to vicinality or miscut, without any layer by layer growth 
oscillations; we neglect considerations of such a step flow growth 
mode in this paper assuming all growth to be occuring strictly on
singular high symmetry substrates.)

The ideal MBE growth (on flat singular high symmetry substrates)
can be thought of as composed of two regimes --- an early time
($t<t_c$) transient (fast diffusion driven) regime of layer by layer
growth followed by the asymptotic ($t>t_c$) kinetically rough
(deposition beam fluctuation driven) growth regime (with no
layer by layer oscillations) characterized by power law
evolution \cite{1,2,3,4,5,6,7,8,9,10,11,12}
in surface roughness.
At low enough temperatures, when surface diffusion is extremely slow,
$t_c$ could be less than the time it takes to grow one monolayer
of deposit on the average, and in that situation layer by layer 
transient growth regime is invisible with the kinetically rough 
growth regime being dominant essentially right from the beginning.
Conversely, sufficiently high temperature growth on a small substrate
could continue in the layer by layer mode for a very long time
although some damping of the growth oscillations is inevitable
with time as distant spatial regions on the substrate must lose 
coherence due to the inherent shot noise fluctuations associated 
with the discrete deposition process in the incident beam.
Thus layer by layer growth us purely a ``finite size'' 
(both spatially and temporally) transient phenomenon ---
if the substrate is made sufficiently large and/or if one waits for
sufficiently long growth time, layer by layer epitaxy must 
necessarily cross over to kinetically rough growth. It is important
to emphasize, however, that the surface diffusion length is 
typically an exponentially activated function of growth temperature,
and therefore a small change in temperature could cause a sharp
large change in the growth mode (layer by layer to rough and vice versa
depending on whether the growth temperature is decreased
or increased) for a given substrate, leading to the empirical 
concept \cite{15} of an epitaxial growth temperature $T_c$ with
growth being layer by layer (rough) for $T>T_c$ ($T<T_c$)
--- clearly $T_c$ is a loosely defined concept because it must be 
a weak (sub-logarithmic) function of the effective substrate
(or the terrace) size for a given material \cite{15}.
In general, ``good'' MBE growth aiming toward producing high
quality smooth epitaxial thin flims is carried out at the highest
possible growth temperature (within the constraint that 
evaporation or desorption from the growth front should be negligible
so that the growth temperature cannot be arbitrarily high)
so as to make atomic mobility at the growth front to be very high
leading to large ``surface diffusion length'' $l$. Here $l$ is 
taken to be the linear size over which the surface is smooth 
due to atomic diffusion.
Assuming the deposition process to be a random Poisson process 
it is then easy to see that the typical surface roughness over
terraces of size $l$ would only grow as 
$\sqrt{\langle h \rangle} /l$, where $\langle h \rangle$
is the thickness of the grown film (and we measure all lengths
in lattice units).
Thus for large $l$, one would have to grow a very thick film 
of thickness $l^2$ before the surface roughness reaches even
one monolayer fluctuation. One can, therefore, grow MBE
thin films of very high smoothness and quality, without worrying
at all about the kinetic surface roughening by properly
adjusting the growth temperature \cite{15} to make $l$ large.

Layer by layer MBE growth has been extensively studied
\cite{16,17,18,19,20}
in the literature using computer simulations of MBE growth through
the stochastic (or kinetic) Monte Carlo simulations, with the
atomistic diffusion at the growth front assumed to be controlled 
by stochastic activated hopping process with the hopping rate 
determined by a local coordination number dependent activated
Arrhenius hopping.
Such activated diffusion Arrhenius hopping simulations 
(sometimes also referred to as ``full diffusion'' simulations
to differentiate them from ``limited mobility'' growth models which
are our main interest in this paper)
involve continuous (possible) hopping of all surface atoms
according to their local bonding configurations (which determine
the activation energy for the hopping process).
Such full diffusion simulations are obviously not well designed to
study the kinetic roughening universality class of MBE growth
because they are extremely time consuming and cannot really be 
carried out for large systems (particularly in the physically
relevant 2+1 dimensions) for long times, an essential requirement
for ascertaining the asymptotic universality class of a growth model.
Although there are some notable exceptions \cite{21,22,23},
the full diffusion Arrhenius activated kinetic Monte Carlo 
simulations of MBE growth have not been used with particular
success for understanding statistical scale invariance properties
of kinetic surface roughening. Instead, important insights into
the MBE universality class of kinetic surface roughening have 
come primarily from nonequilibrium limited mobility growth models
--- mainly the so-called Wolf-Villain \cite{11} (WV) and the
Das Sarma-Tamborenea \cite{10} (DT) model --- which were
introduced specifically for the elucidation of the MBE growth 
universality. 

In this paper we study the DT and the WV model (we emphasize that
WV and DT models, in spite of their close similarity in growth
rules, belong \cite{24} to different asymptotic universality
classes in both 1+1 and 2+1 dimensions although their pre-asymptotic
scaling behavior is very similar which has led to considerable
confusion in the literature) 
in the complementary layer by layer growth regime rather than
the kinetically rough growth regime which motivated the introduction
of these models. We mention in this context that some 1+1
dimensional studies of WV and DT models in the layer by layer 
growth regime have recently been reported in the literature
\cite{25,26}. Our results, where applicable, agree with these
earlier works \cite{25,26}, but our focus in this paper is 2+1 
dimensional growth and the effect of long surface diffusion length
in 1+1 dimensional growth, neither one of which has 
earlier been studied.

In limited mobility growth models (the models and growth rules
used in this paper are described in section II of the paper
--- see, for example, Fig. \ref{f1}),
in sharp contrast to full diffusion MBE growth simulations,
the goal is to suppress crossover and transient effects as
much as possible (so as to efficiently reach the asymptotic 
kinetic surface roughening regime)
and as such only the most recently deposited atom is allowed
to diffuse or relax instantaneously to the appropriate incorporation
site following the mobility rules of the specific model.
This allows suppression of crossover effects invariably
present in the full diffusion simulations arising from many 
different diffusion rates corresponding to many different 
possibilities for local bonding configurations --- the only
time scale in the limited mobility growth models being the
deposition rate, which defines the time unit for the problem.
From now on we take the time unit (sometimes referred to as a
``second'') as the time to deposit one monolayer on the average.
Thus the growth time in this paper also defines $\langle h \rangle$
--- the average thickness of the deposited film measured in units
of monolayers or the lattice constant, which we take to be the
unit of length throughout. (With no loss of generality we
choose the lattice constant to be the same along the substrate 
and the growth directions.)

The limited mobility growth models \cite{10,11} are by construction
strongly dominated by the deposition shot noise because the goal
is to study the scale invariant kinetic surface roughening behavior.
This is particularly true in the original versions of the growth
model where the surface diffusion length is choosen to be unity,
$l=1$, i.e. the deposited atoms are allowed to move only to the
nearest neighbor incorporation sites around the deposition site.
The original DT and WV models therefore did not exhibit, by design,
any layer by layer growth oscillations since the smoothing or
the healing distance ($l$) was just one lattice unit.
In order to manifest layer by layer epitaxy in limited mobility 
growth models one must therefore suppress the shot noise 
associated with the incident beam fluctuations.

In this paper we accomplish the noise suppression by two alternative
techniques : The `multiple hit' noise reduction technique 
\cite{27,28,29,30,31}
and the `long surface diffusion length ($l>1$)' noise reduction
technique \cite{21}.
These techniques, described in section II of the paper,
give rise to layer by layer growth (as monitored by an
oscillatory surface roughness, i.e. $W(t)$ showing oscillations
as a function of growth time $t$)
in the limited mobility growth models as described in
section III of the paper.

The rest of this paper is organized as follows. In section II
we describe the limited mobility growth models and the noise
reduction technique(s) employed by us.
We also provide some theoretical background for our analysis of 
the simulation results. In section III we present and describe
our numerical simulation results for layer by layer epitaxial
growth in DT and WV models. We also discuss in section III 
various (approximate) scaling properties of our simulated
layer by layer epitaxial growth results. We conclude in section
IV with a general discussion of our results making connections
with some of the existing results in the literature and
pointing out possible future directions as well.

\section{Models, Theory, and Background}

The DT \cite{10} and WV \cite{11} models used in our simulations are
shown in Fig. \ref{f1}. We carry out growth simulations in both 1+1
dimensions and 2+1 dimensions (on 100 high symmetry substrates).
Particles are dropped, obeying solid on solid constraint, 
sequentially (one by one) and randomly with an average rate of
$N$ per second, where $N=L$ for d=1+1 and $N=L^2$ for d=2+1,
on a substrate of lateral linear size of $L$ lattice units.
We measure length in lattice units and time in inverse deposition
rate (i.e. in units of monolayer filling since one average 
monolayer of deposition occures every ``second'').
Each deposited atom is allowed to ``diffuse'' instantaneously to 
its incorporation site following the mobility rules of the specific
model. The diffusion rules in the DT model are that a deposited
atom can move only if it has no lateral nearest neighbor in the
same layer (if it does, then the atom is incorporated at the
deposition site) --- if the deposition site has no lateral
nearest neighbor then the incident atom may move instantaneously
to a neighboring empty site (within a lateral diffusion length
of $l$, where $l=1$ in the original model and in most existing 
simulations) provided the final incorporation site has a higher 
lateral coordination number (i.e. one or higher) than the deposition
site. If several neighboring sites satisfy the diffusion rule
then the atom will move randomly to any one of them with equal 
probability. The rules for the WV model are superficially
similar to the DT rules : In the WV model all deposited atoms
(and {\it not} just the ones with no lateral bonds) can, in principle,
move provided they can increase their local lateral bonding and
the deposited atom always moves to the site with the maximum 
local bonding environment. In both models, the deposited atom is
incorporated at the deposition site if it cannot satisfy the
diffusion rules (i.e. no sites with higher coordination available)
within the diffusion length.

Both DT and WV models have been extensively studied in the literature
(mostly within nearest neighbor, $l=1$, diffusion rules) in the 
context of their kinetic surface roughening universality classes.
Recently, layer by layer epitaxy in the WV \cite{25,26,30}
and the DT \cite{25} model have been investigated in 1+1 dimensions
using the multiple hit noise reduction technique.
The very first simulational observation of layer by layer growth
in a limited mobility growth model was reported in the DT model
in Ref. \cite{21} where it was studied in 1+1 dimensions using
a long ($l>1$) surface diffusion length, but no details were
investigated. We emphasize that the usual $l=1$ limited mobility
growth model does {\it not} exhibit any layer by layer epitaxy
by definition, and manifests kinetic roughening right from the
beginning since for $l=1$ the layer by layer epitaxy regime is
restricted to less than one monolayer coverage, i.e. in the 
standard limited mobility growth models \cite{10,11,12}
the layer by layer epitaxy regime does not exist.

To obtain a layer by layer growth regime in the DT and the WV model
we use two distinct techniques to suppress noise and enhance diffusion,
which enable our growth simulations to manifest strong layer by
layer growth oscillations before crossing over to the kinetic surface
roughening regime with pure multilayer growth. One technique, 
referred to as the noise reduction technique, has also been used
by others \cite{25,26,27,28,29,30}
to produce layer by layer growth in various models, e.g.
Eden model \cite{28}, single step model \cite{29}, and the WV 
and the DT model in 1+1 dimensions \cite{25,26,30}.
We have earlier used this technique \cite{31} to suppress corrections
to scaling in the asymptotic kinetically rough surface growth regime
in the DT and the WV model in order to accurately determine the
dynamic scaling exponents and the associated growth universality
class. In the noise reduction technique, characterized by an integer
number $m$ (the usual growth model without any noise reduction is 
an $m=1$ model), a counter is put on each surface site and each
discrete deposition event on a site advances the counter by unity.
A deposition event at a particular site is accepted only when the 
counter reaches a pre-determined number $m>1$. Thus, this 
technique is the multiple hit noise reduction technique since
$m(>1)$ deposition hits on a site are needed for a true deposition.
After $m$ hits on a site (i.e. after the acceptance of a deposition
event) the counter at that particular site is set back to zero,
and the whole multiple hit process begins all over again.
The multiple hit noise reduction technique is a coarse-graining
procedure which suppresses the deposition shot noise, and the
noise reduction is enhanced for larger values of $m$.
The second technique applied by us for obtaining layer by layer
growth oscillations in limited mobility growth models is to use
long surface diffusion lengths ($l>1$) in the growth simulations.
Obviously long diffusion lengths enhance the layer by layer growth
regime, and in particular, for $l>L$ i.e. the surface diffusion
length exceeding the system size, the layer by layer growth may
persist essentially indefinitely since each deposited atom may
always be able to seek out a desired epitaxial site for 
incorporation. In some sense the multiple hit noise reduction
parameter $m$ is equivalent to the dimensionless diffusion
length parameter $l/L$ because large values of both tend to
enhance the layer by layer growth regime. In section III, 
where we present our simulation results, we will see the precise
nature of this correspondence between $m$ and $l/L$ in our
two methods of obtaining layer by layer epitaxy in the DT
and the WV model.

The central quantity of interest in layer by layer epitaxial 
growth is the characteristic time $t_c$ at which layer by layer
growth dies out, i.e. for the deposited average film thickness
larger than $\langle t_c \rangle$, measured in lattice units
or in monolayers, there are no discernible layer by layer
growth oscillations. It has been found in earlier numerical
simulations of layer by layer growth in a variety of contexts
that $t_c$ obeys an approximate scaling relation with the
coarse graining parameter $m$ (or $l/L$ as the case may be),
and we are interested in investigating whether the following 
scaling relations hold in the limited mobility growth models
\begin{equation}
t_c \sim \left \{ \begin{array}{ll}
          m^\mu         & \mbox{for the noise reduction method} \\
         (l/L)^\delta  & \mbox{for the long diffusion length method.}
                \end{array}
       \right.
\label{e1}
\end{equation}
If such scaling relations do hold in our simulations we are interested 
in obtaining the
relationship, if any, between the exponents $\delta$ and $\mu$.

It is in fact fairly straightforward to obtain a relationship 
between the noise reduction
parameter $m$ and the surface diffusion length parameter $l/L$ 
using only dimensional
arguments. In particular, we note that for a $d^\prime$-dimensional 
substrate ($d^\prime=2$ for real
surfaces and $d^\prime=1$ for the 1+1 -dimensional growth) one obtains 
a simple relationship
between $\mu$ and $\delta$ by noting that a surface diffusion 
length $l$ corresponds to 
$m \sim l^{d^\prime}$ since there are 
$l^{d^\prime}$ available sites for 
a particular deposited 
particle to incorporate at. This immediately leads to
\begin{equation}
\delta = \mu d^\prime.
\label{e2}
\end{equation}
Within the limited accuracy of our growth simulations 
(see section III for the results),
we find the scaling relation defined by Eq. (\ref{e2}) to be valid.

One secondary theoretical goal of our work is to investigate the
extent to which layer by layer growth oscillations obey scaling 
with respect to growth time (or equivalently, the average film
thickness). Using Eq. (\ref{e1}) as the theoretical ansatz one can 
ask whether the surface roughness $W$, defined as the ensemble 
averaged (over many growth simulations) root mean square 
fluctuation in the interface height, which is oscillatory
(and $W<1$ implying little roughness) in the layer by layer growth
regime is a general scaling function of the layer by layer growth
parameter $m$ or $l/L$ through a dependence of the form :
\begin{equation}
W(t) \sim f_m(t/m^\mu) \mbox{\,\, or \,\, } f_l(t/(l/L)^\delta).
\label{e3}
\end{equation}
If a scaling form such as Eq. (\ref{e3}) holds in the layer by layer 
growth regime, then Eq. (\ref{e1}) for $t_c$ trivially follows from
it --- $t_c$ being the value of time where layer by layer 
oscillations cease to exist. We could go further in our scaling 
analyses and ask whether the scaling defined by Eq. (\ref{e3})
continues to hold (perhaps approximately) well beyond 
($t>t_c$) the layer by layer growth regime establishing an 
approximate scaling relationship between the layer by layer growth
regime ($t<t_c$) and the kinetically rough growth regime
($t>t_c$). Our results presented in section III indicate that 
such an approximate scaling relation does indeed exist between
the layer by layer growth regime and the kinetically rough
growth regime.

Finally, we note that there have been recent attempts
\cite{32,33} at developing a theory for layer by layer growth 
oscillations starting from continuum growth equations underlying
the coarse grained long wavelength behavior of MBE growth.
A simple dimensional argument \cite{32}, later followed up \cite{33}
by a renormalization group approach, leads to the following results
for MBE growth :
\begin{equation}
\mu = \left \{ \begin{array}{ll}
          4/3  & \mbox{for $d^\prime=1$} \\
          2    & \mbox{for $d^\prime=2$.}
               \end{array}
      \right.
\label{e4}
\end{equation}
The exponents defined by Eq. (\ref{e4}) correspond to the 
so-called \cite{1,2,3,4,5,6} conserved fourth order nonlinear
growth equation or the nonlinear MBE growth equation (which is
sometimes also referred to as the conserved KPZ equation with
nonconserved noise or the Lai-Das Sarma-Villain \cite{7,8},
LDV, equation) which is given by
\begin{equation}
\frac{\partial h}{\partial t} = - \nu_4 \nabla^4 h
   + \lambda_{22} \nabla^2 (\mbox{\boldmath $\nabla$} h)^2
   + \eta,
\label{e5}
\end{equation}
where $h(\bf{x},t)$ is the dynamical height fluctuation variable
relative to the average interface $\langle h \rangle$ at the 
substrate site $\bf{x}$ (with $\bf{x}$ is the lateral substrate
coordinate), 
$\mbox{\boldmath $\nabla$} \equiv \frac{\partial}{\partial \bf{x}}$
is the gradient operator along the surface, 
$\eta$ is the deposition shot noise (which causes the kinetic 
surface roughening), and $\nu_4$, $\lambda_{22}$ are coefficients
which in general depend on surface diffusion rate, 
deposition rate, etc. Since the continuum description of the DT and 
the WV model are actually quite complex \cite{23,31,34,35,36,37}, 
and are likely to be different in different dimensions \cite{37},
it is by no means clear that the exponents (given in Eq. (\ref{e4}))
corresponding to Eq. (\ref{e5}) apply without any qualifications 
to the DT and the WV model (as will be discussed in section III 
where we present our numerical simulations).
We therefore also provide below the exponent $\mu$ for the linear
second order Edwards-Wilkinson (EW) growth equation \cite{38}
which applies to the limited mobility Family (F) growth model
\cite{39} and may also have significant relevance to the DT
and the WV model \cite{31,37,40} :
\begin{equation}
\mu = \left \{ \begin{array}{ll}
          2         & \mbox{for $d^\prime=1$} \\
          \infty    & \mbox{for $d^\prime=2$.}
               \end{array}
      \right.
\label{e6}
\end{equation} 
The EW equation, whose layer by layer growth exponent in 1+1 and
2+1 dimensions is given in Eq. (\ref{e6}), is the following :
\begin{equation}
\frac{\partial h}{\partial t} = \nu_2 \nabla^2 h + \eta.
\label{e7}
\end{equation}
We note here that both sets of exponent values given by 
Eq. (\ref{e4}) and (\ref{e6}) will be relevant in our discussion
of our simulation results to be presented in the next section.

\section{Results and Discussions}

We now present our 1+1 (i.e. $d^\prime=1$) and 2+1 (i.e. $d^\prime=2$)
dimensional layer by layer growth simulation results for the discrete
limited mobility DT and WV models in Figs. \ref{f2}-\ref{f6}.
In Figs. \ref{f2}-\ref{f4} we present 1+1 dimensional simulation
results whereas Figs. \ref{f5} and \ref{f6} give 2+1 dimensional
simulational results for the two growth models. In each figure
(to be described below) the panel (a) gives the simulated $W(t)$
as a function of growth time $t$ for various values of the noise
reduction parameter $m$ or the surface diffusion length $l$
with the layer by layer oscillations manifestly obvious for larger
values of $m$ and $l$.
(The original DT and WV models correspond to the simulation with 
$m=1$ and $l=1$, which has no layer by layer oscillations 
by construction.)
In panel (b) of each figure we demonstrate our best computed 
scaling collapse of the $W-t$ plots (shown in panel (a))
for various values of $m$ or $l$ with a suitable scaling of time
$t$ to $t/m^\mu$ or $t/(l/L)^\delta$ as the case may be.
For each scaling collapse in (b) we try various different values
of the exponent $\mu$ or $\delta$ ro obtain the best statistical
scaling in the simulated data. The system size used in each 
simulation is indicated in the corresponding figures and the captions.
Here we should emphasize that the results shown in this paper
represent only a typical fraction of our extensive DT
and WV layer by layer growth simulations. The representative 
results presented here are of course in complete agreement with
the full set of our simulation data, and our conclusion is based
on a very large set of simulation results and not just on the 
results presented in this paper.

In Fig. \ref{f2} we show our 1+1 dimensional DT layer by layer growth
simulation results for finite surface diffusion length $l$
($m=1$, $l>1$) for $l=10,\,20,\,30,\,40,\,50$. The layer by layer
oscillations are visually obvious in Fig. \ref{f2}(a) ---
the damping time $t_c$ increases from $t_c \sim 10$ for $l=10$ 
to roughly $t_c \sim 50$ for $l=50$.
In general, the magnitude of the oscillations decays exponentially 
with increasing time, and for $t>t_c$ we can only discern the
power law increase of $W \sim t^\beta$ where $\beta$ is 
the growth exponent in the model. In Fig. \ref{f2}(b) we show 
our scaling collapse of the $W(t)$ data from Fig. \ref{f2}(a),
leading to the exponent value $\delta \approx 1.5$.
Thus, $t_c \sim (l/L)^{1.5}$ in $d^\prime = 1$ DT model.
We note that the scaling collapse in the kinetically rough 
growth regime (i.e. for $t>t_c$) is not excellent, but it is 
remarkable that the exponent $\delta$ which is meaningfully
defined only in the layer by layer growth regime continues to 
provide an approximate reasonable description of the kinetically
rough growth regime.

In Fig. \ref{f3} we show our 1+1 dimensional DT model layer by
layer growth oscillations (Fig. \ref{f3}(a)) for the multiple hit
noise reduction technique ($l=1$, $m>1$) for different values of
the noise reduction factor $m$. In Fig. \ref{f3}(b) we show the
scaling collapse of the $W-t$ data in Fig. \ref{f3}(a) for
various $m$ values. The scaling is excellent with an exponent
$\mu = 1.5$. Thus, $t_c \sim m^{1.5}$ in $d^\prime = 1$ DT model.

In Fig. \ref{f4} we depict our noise reduced layer by layer growth
oscillations (Fig. \ref{f4}(a)) and the corresponding scaling 
collapse of the data for various values of $m$ (with $l=1$) in
the 1+1 dimensional WV model simulations. Again, the scaling exponent
$\mu$ is found to be $\mu=1.5$ for the best scaling collapse,
indicating $t_c \sim m^{1.5}$ in both DT and WV noise reduced models
in $d^\prime=1$. 
The finding of the apparent same exponent value $\mu=1.5$ in both DT 
and WV models in 1+1 dimensional growth is consistent with the fact
that the effective growth exponent $\beta$ (obtained by plotting
$\ln \, W$ against $ln \, t$ in the simulated results) is 
almost identical in the two models :
From the slope of the $\log-\log$ plot in Fig. \ref{f2}(b) we obtain
$\beta \approx 0.338$ for the $d^\prime=1$ DT model whereas from
the slope of the $\log-\log$ plot in Fig. \ref{f3}(b) we obtain
$\beta \approx 0.339$ for the $d^\prime=1$ WV model. 
Thus within the effective time and length scales of our simulations
the two models (DT and WV) have essentially the same effective
dynamical universality class, which is consistent with the fact that
they  have the same effective exponent $\mu=1.5$ in $d^\prime=1$.
The fact that the asymptotic universality classes of the DT and WV
models are different even in $d^\prime=1$ dimension 
\cite{31,37,40} does not seem to affect the effective values of 
$\mu$ we obtain in our simulations.

Before presenting our 2+1 dimensional simulation results in
Fig. \ref{f5}-\ref{f7} we first discuss the exponent values 
$\delta$ and $\mu$, all of which have turned out to be approximately 
$1.5$ in the 1+1 dimensional DT and WV layer by layer epitaxial 
simulations. (We do not show here $l>1$, $m=1$ simulation results
for the $d^\prime=1$ WV model because they are very similar to 
those shown in Figs. \ref{f2}-\ref{f4} with the same exponent
$\mu \approx 1.5$.)
First we note that the exponent value is very close to 
(but somewhat above) the theoretically ``expected'' exponent
$\mu = 4/3 = 1.33$ predicted in Refs. \cite{32,33}
assuming that the continuum growth equation for these discrete
growth models is that given in Eq. (\ref{e5}). We also note that
the expected relationship between $\mu$ and $\delta$, namely
$\delta = \mu d^\prime$ which becomes $\delta = \mu$ 
in $d^\prime=1$, is obeyed by our simulation results. 
We also add that the measured exponent $\mu \simeq 1.5$ is
consistent with other findings in the literature
\cite{25,26,30}. The cause for our calculated $\mu \, (\simeq1.5)$ 
to be somewhat (by roughly $10 \%$) higher than the theoretical
value ($\mu = 4/3$) is not very clear at this stage.
We do not, however, believe this discrepancy to be particularly
significant because of a number of reasons :
(1) our scaling collapse are in fact not inconsistent with an
exponent of $1.33$ although an exponent value of $1.5$ is 
definitely a statistically better fit for our scaling collapse;
(2) it is quite conceivable that there are some systematic 
finite size and finite time corrections to scaling 
(which are known to be very important in DT and WV models);
(3) finally, at least in the WV model which is definitely 
known \cite{31,37,40} to asymptotically belong to the EW
universality class, it is possible that our simulated exponent
$\mu \simeq 1.5$ is showing some effects of the asymptotic 
universality class since the theoretically expected
\cite{32,33} $\mu$ for the linear EW equation 
(our Eq. (\ref{e7})) is $\mu=2$ in $d^\prime=1$ dimensions.

Our $d^\prime=2$ dimensional noise reduced ($m>1$, $l=1$)
results for the DT, the WV, and the F model are shown in 
Figs. \ref{f5}-\ref{f7} respectively. These 2+1 dimensional 
layer by layer growth results (as well as the results shown in
Fig. \ref{f2}) in limited mobility models are completely
new and do not exist anywhere in the literature.
We carried out our 2+1 layer by layer growth simulations using
only the noise reduction technique since our 1+1 dimensional 
results (compare Figs. \ref{f2} and \ref{f3}) indicate that the
finite diffusion length and the multiple hit noise reduction 
techniques are essentially equivalent and the 2+1 dimensional
simulations with $l>1$ are particularly cumbersome to carry out.
Our $d^\prime=2$ layer by layer epitaxy simulations seem to 
have produced a few surprising results as discussed below.

In Fig. \ref{f5} we present our $d^\prime=2$ layer by layer 
growth simulations in the DT model using the noise reduction
technique. The results are depicted in the same manner as in
the $d^\prime=1$ case shown in Fig. \ref{f3} --- in particular,
Fig. \ref{f5}(a) shows the actual layer by layer growth 
oscillations for various values of $m$ whereas Fig. \ref{f5}(b)
shows the scaling collapse. It is obvious from comparing 
Fig. \ref{f3} and \ref{f5} that the layer by layer growth regime
is substantially stronger in $d^\prime=2$ case compared with the
$d^\prime=1$ case, which is consistent with a much larger value
of the damping exponent $\mu \simeq 2.5$ (compared with $1.5$
in $d^\prime=1$) in Fig. \ref{f5}(b) in the $d^\prime=2$ system.
Our calculated damping exponent $\mu \simeq 2.5$ for the 
$d^\prime=2$ dimensional DT model is substabtially (by $25 \%$)
higher than the corresponding theoretical prediction 
\cite{32,33} of $\mu=2$ for the LDV equation (Eq.(\ref{e5})),
which is generally thought to be the continuum description
for MBE growth. This large discrepancy between our simulated
damping exponent ($\mu \simeq 2.5$) and the theoretical
damping exponent ($\mu = 2$) corresponding to Eq. (\ref{e5}) in
$d^\prime=2$ dimensions may be a real effect, arising from the
recently discovered fact \cite{41} that the DT model in 2+1
dimensions has actually a very small (but non-zero) EW
$\nabla^2 h$ term in its continuum description in contrast 
to the 1+1 dimensional DT model where the EW $\nabla^2 h$ 
term strictly vanishes (i.e. $\nu_2 = 0$ in Eq. (\ref{e7}))
by virtue of a topological symmetry \cite{6} in the DT model.
Thus the DT model in 2+1 dimensions may actually asymptotically
belong (but with a very small value of $\nu_2$) to the EW
universality class which according to Eq. (\ref{e6}) has an
infinite value of the damping exponent $\mu$.
It is therefore possible that the value $\mu \simeq 2.5$ in
Fig. \ref{f5}(b) may be indicative of a small correction in
the 2+1 dimensional DT model arising from the EW term in the
continuum equation. More work is needed to conclusively settle 
this issue. This would also explain why the scaling collapse 
for the damped oscillations in Fig. \ref{f5}(b), particularly 
for the data in the kinetically rough ($t>t_c$) growth regime,
is not as good as the corresponding 1+1 dimensional results shown
in Fig. \ref{f3} --- the asymptotic corrections to the LDV
equation (Eq. (\ref{e5})) arising from a small $\nabla^2 h$
term (which is present in $d^\prime=2$ DT simulation results 
of Fig. \ref{f5}.

The noise reduced $d^\prime=2$ WV model simulations, shown in 
Fig. \ref{f6}, are even more surprising. The layer by layer growth
oscillations in the initial transient time (upto $t \sim 10$ or so)
are apparent in Fig. \ref{f6}(a) for finite values of $m$ although
the oscillations are already weaker than the corresponding DT
results shown in Fig. \ref{f5}. This is not expected based on 
the $d^\prime=1$ results where Fig. \ref{f3} (DT) and Fig. \ref{f4}
(WV) are essentially identical within our simulation sizes and times.
Even more surprising is the scaling collapse in the $d^\prime=2$ WV 
model shown in Fig. \ref{f6}(b) which is clearly not a good scaling
behavior --- in fact, beyond the layer by layer regime
(i.e. for $t>t_c$) there is essentially no scaling behavior in the
$d^\prime=2$ WV results. The scaling behavior of the WV model
(Fig. \ref{f6}) in the 2+1 dimensional WV model is clearly very
different from (and worse than) the corresponding DT results.
The best damping exponent value for $\mu$ we obtain from
Fig. \ref{f6}(b) is $\mu \approx 1.5$, which is the same as 
the corresponding WV value in $d^\prime=1$ as depicted in 
Fig. \ref{f4}. We should emphasize that this estimate
($\mu \approx 1.5$) for the damping exponent in the 2+1 dimensional
WV model should be taken at best as a crude estimate for an
effective exponent since there is no scaling behavior in the WV
data shown in Fig. \ref{f6}. 
Given the very similar $d^\prime=1$ behavior in the DT and WV models
(as shown in Figs. \ref{f3} and \ref{f4}) it is very surprising 
that the $d^\prime=2$ behavior in the two models (including the
effective damping exponent values $\mu \simeq 2.5$ and $1.5$
respectively for the DT and the WV model) is so completely different.
 
What is the explanation for this striking difference in the
$d^\prime=2$ layer by layer epitaxial growth behavior in the DT
(Fig. \ref{f5}) and the WV (Fig. \ref{f6}) model (particularly in view
of their essentially identical behavior in $d^\prime=1$ as seen in
Figs. \ref{f3} and \ref{f4})?
The explanation actually lies in the recently discovered fact
\cite{41,42} that, while the $d^\prime=1$ dimensional WV model obeys
\cite{37,40,41} the continuum growth equations given in Eqs. (\ref{e5})
and (\ref{e7}) with $\nu_4$ and $\lambda_{22} \neq 0$ and
$\nu_2$ very small but having a nonvanishing positive value, 
the $d^\prime=2$ dimensional WV growth model is {\it actually 
unstable} in the sense that the WV morphology in the 2+1 dimensional 
growth forms a regular mounded structure with the mound edges
having approximately constant slopes. Such an epitaxial mounding 
instability \cite{42} in the 2+1 dimensional WV model becomes
particularly manifest under the noise reduction technique as
discussed in details in Ref. \cite{42}. This unstable mounded
morphological growth \cite{42} in the noise reduced $d^\prime=2$
WV model leads to the peculiar behavior seen for late times in
Fig. \ref{f6}, where the epitaxial mounding instability prevents 
the usual layer by layer growth regime from behaving in the
``usual'' manner depicted in Figs. \ref{f2}-\ref{f5}.

Finally, in Fig. \ref{f7} we show our 2+1 dimensional Family (F)
model simulations under noise reduced ($m>1$, $l=1$) conditions.
The F model, which by construction is designed to follow exactly
the linear second order Edwards-Wilkinson equation (Eq. (\ref{e7}))
with $\nu_4=\lambda_{22}=0$ in Eq. (\ref{e5}) and $\nu_2 \neq 0$,
has very simple growth rules (not shown in Fig. \ref{f1}) :
Each randomly deposited atom seeks to find the site of local 
height minina as the incorporation site. The F model is essentially
the discretized version of the EW equation (Eq. (\ref{e7})), 
and as such has the EW dynamical exponents : $\beta=0$
and $\mu=\infty$ in $d^\prime=2$. The growth exponent $\beta$
(defines the presaturation kinetic roughening of $W$ as
$W \sim t^\beta$ \cite{1,2,3,4,5,6}) being zero (i.e. $W \sim
\ln \, t$) in the F model, growth is already very smooth 
because the roughening is only logarithmic in time.
In the presence of noise reduction, therefore, the F model
(Fig. \ref{f7}(a)) shows persistent layer by layer growth
oscillations in $d^\prime=2$ with only logarithmic damping of the
oscillation induced by kinetic surface roughening.
Since the damping exponent $\mu$ in the noise reduced $d^\prime=2$
F model is infinity (Eq. (\ref{e6})), we cannot obtain any scaling 
collapse of the layer by layer growth simulation data of Fig.
\ref{f7}(a) which is obvious in the ``scaling plot'' shown in 
Fig. \ref{f7}(b). A very large value of the exponent $\mu$
($\approx 100$, for example) will of course produce trivial 
(and meaningless) data collapse, but we have checked that no
finite reasonably small (upto $\mu = 10$) value of the damping 
exponent $\mu$ produces scaling in Fg. \ref{f7}(b).
Thus, our $d^\prime=2$ F results are consistent with the theoretical
prediction \cite{32,33} of $\mu$ being infinity in the $d^\prime=2$
EW equation. We note that we have also carried out $d^\prime=1$
dimensional noise reduced layer by layer growth simulations
(not presented in this paper) in the F model, obtaining excellent
scaling collapse with the theoretically predicted value of 
$\mu = 2$. Our results for the $d^\prime=1$ F model layer by layer
growth are consistent with those reported in Ref. \cite{43}.

\section{Conclusion}

We have presented numerical results for extensive computer simulations
of various noise reduced limited mobility DT \cite{10}, WV \cite{11},
and F \cite{39} growth models in 1+1 and 2+1 dimensions in order to
study the damping of the layer by layer growth epitxy invariably 
induced by the shot noise inherent in the deposition beam fluctuations.
We have used both multiple hit noise reduction technique 
and the long surface diffusion length method to obtain the layer by 
layer growth in our simulations, and have shown that these two
different techniques for obtaining layer by layer growth are
essentially equivalent. Our simulation results in general exhibit
(with two exceptions noted below) very good scaling connecting the
layer by layer growth regime with the kinetically rough growth regime.
Our calculated damping exponents agree well with theoretical 
predictions where applicable. The two exceptions noted above are the
2+1 dimensional WV and the F model where scaling fails for different
reasons. The 2+1 dimensional noise reduced WV model is known 
\cite{42} to manifest unstable growth with spectacular epitaxial
mounding, which inhibits layer by layer growth leading to the
failure of scaling collapse. The 2+1 dimensional F model on the
other hand exhibits very strong and persistent layer by layer growth
oscillations (with little kinetic roughening) whose damping is
expected on theoretical grounds to be extremely weak leading to
an infinite value of the damping exponent, which is equivalent to
saying that there is essentially no scaling since in the presence 
of noise reduction layer by layer growth regime lasts forever.

This work has been supported by the US-ONR.

\begin{figure} \caption{
Schematic plots showing the diffusion rules for the DT and
WV models when the diffusion length (l) and noise reduction
factor (m) are both one.
\label{f1}} \end{figure}

\begin{figure} \caption{
(a) W-t oscillations for 1+1 DT ($L=1000$) with m=1
and l=10, 20, 30, 40, 50 (top to bottom),
(b) scaling plot of systems in (a) using $\delta=1.5$. 
\label{f2}} \end{figure} 

\begin{figure} \caption{
(a) W-t oscillations for 1+1 DT ($L=1000$) with l=1
and m=1, 5, 8, 10, 15 (top to bottom),
(b) scaling plot of systems in (a) using $\mu=1.5$.
\label{f3}} \end{figure} 

\begin{figure} \caption{
(a) W-t oscillations for 1+1 WV ($L=1000$) with l=1
and m=1, 5, 8, 10, 15 (top to bottom),
(b) scaling plot of systems in (a) using $\mu=1.5$.
\label{f4}} \end{figure} 

\begin{figure} \caption{
(a) W-t oscillations for 2+1 DT ($L=10^3 \times 10^3$) 
with l=1 and m=1, 5, 8, 10, 15 (top to bottom),
(b) scaling plot of systems in (a) using $\mu=2.5$.
\label{f5}} \end{figure} 

\begin{figure} \caption{
(a) W-t oscillations for 2+1 WV ($L=100 \times 100$)
with l=1 and m=1, 5, 8, 10, 15 (top to bottom),
(b) scaling plot of systems in (a) using $\mu=1.5$.
\label{f6}} \end{figure} 

\begin{figure} \caption{
(a) W-t oscillations for 2+1 Family ($L=100 \times 100$)
with l=1 and m=1, 5, 8, 10 (top to bottom),
(b) scaling plot of systems in (a) using $\mu=0.0$.
\label{f7}} \end{figure}


\begin{thebibliography}{99}
%
\bibitem{1}
T. Vicsek, {\it Fractal Growth Phenomena} (World Scientific Singapore,
1992).
%
\bibitem{2}
A. L. Barabasi and H. E. Stanley, {\it Fractal Concepts in Surface
Growth} (Cambridge, New York, 1995).
%
\bibitem{3}
P. Meakin, {\it Fractals, Scaling and Growth Far From Equilibrium},
(Cambridge, New York, 1998).
%
\bibitem{4}
A. Pimpinelli and J. Villain, {\it Physics of Crystal Growth},
(Cambridge, New York, 1998).
%
\bibitem{5}
T. Halpin Healy and Y. C. Zhang, Phys. Rep. {\bf 254}, 215 (1995).
%
\bibitem{6}
J. Krug, Adv. Phys. {\bf 46}, 139 (1997).
%
\bibitem{7}
Z. W. Lai and S. Das Sarma, Phys. Rev. Lett. {\bf 66}, 2348 (1991).
%
\bibitem{8}
J. Villain, J. Phys. I {\bf 1}, 19 (1991).
%
\bibitem{9}
M. Kardar, G. Parisi, and Y. C. Zhang, Phys. Rev. Lett. {\bf 56},
889 (1986).
%
\bibitem{10}
S. Das Sarma and P. I. Tamborenea, Phys. Rev. Lett. {\bf 66}, 325 (1991).
%
\bibitem{11}
D. Wolf and J. Villain, Europhys. Lett. {\bf 13}, 389 (1990).
%
\bibitem{12}
S. Das Sarma and S. V. Ghaisas, Phys. Rev. Lett. {\bf 69}, 3762 (1992).
%
\bibitem{13}
J. M. van Hove et. al., 
J. Vac. Sci. Tech. B {\bf 1}, 741 (1983);
P. I. Cohen et. al., Surf. Sci. {\bf 216}, 222 (1989);
G. S. Petrich et. al.,J. Crys. Growth {\bf 95}, 23 (1989).
%
\bibitem{14}
P. K. Larsen and P. J. Dobson, eds., {\it Reflection High-Energy
Electron Diffraction and Reflection Imaging of Surfaces}
(Platinum, New York, 1988);
M. G. Lagally, ed., {\it Kinetics of Ordering and Growth at Surface}
(Plenum, New York, 1990).
%
\bibitem{15} S. Das Sarma and P. I. Tamborenea, Phys. Rev. B
{\bf 46}, 1925 (1992).
%
\bibitem{16}
S. Clarke and D. D. Vvedensky, Phys. Rev. Lett. {\bf 58}, 2235 (1987);
J. Appl. Phys. {\bf 63}, 2272 (1988);
S. Clarke, M. R. Wilby and D. D. Vvedensky, Surf. Sci. 
{\bf 255}, 91 (1991); 
S. A. Barnett and A. Rockett, Surf. Sci. {\bf 198}, 133 (1988).
%
\bibitem{17}
T. Kawamura, A. Kobayashi and S. Das Sarma, Phys. Rev. B {\bf 39},
12723 (1989).
%
\bibitem{18}
S. Das Sarma, S. M. Paik, K. E. Khor and A. Kobayashi, J. Vac. Sci.
Tech. B {\bf 5}, 1179 (1987).
%
\bibitem{19}
I. K. Marmorkos and S. Das Sarma, Surf. Sci. Lett. {\bf 237},
L411 (1990); Phys. Rev. B {\bf 45}, 11262 (1992).
%
\bibitem{20}
S. Das Sarma, J. Vac. Sci. Tech. B {\bf 10}, 1695 (1992).
%
\bibitem{21}
P. I. Tamborenea and S. Das Sarma, Phys. Rev. E {\bf 48}, 2575 (1993).
%
\bibitem{22} M. R. Wilby, D. D. Vvedensky and A. Zangwill, 
Phys. Rev. E {\bf 48}, 852 (1993);
M. Kotrla and P. Smilaner, Phys. Rev. B {\bf 53}, 13777 (1996);
S. Pal and D. P. Landau, Phys. Rev. B {\bf 49}, 10597 (1994).
%
\bibitem{23}
S. Das Sarma, C. J. Lanczycki, R. Kotlyar and S. V. Ghaisas,
Phys. Rev. E {\bf 53}, 359 (1996).
%
\bibitem{24}
P. Punyindu, Ph.D. thesis (department of Physics, University of 
Maryland, 2001; unpublished).
%
\bibitem{25}
L. Brendel, H. Kallabis and D. E. Wolf, Phys. Rev. E {\bf 58},
664 (1998).
%
\bibitem{26}
S. Park, H. Jeong and B. Kahng, Phys. Rev. E {\bf 59}, 6184 (1999).
%
\bibitem{27}
J. Kertesz and T. Vicsek, J. Phys. A {\bf 19}, L257 (1986).
%
\bibitem{28}
D. E. Wolf and J. Kertesz, Europhys. Lett. {\bf 4}, 651 (1987);
J. Kertesz and D. E. Wolf, J. Phys. A {\bf 21}, 747 (1988).
%
\bibitem{29}
L. H. Tang, p.99 in {\it Growth Patterns in Physical Sciences 
and Biology}, edited by J. M. Garcia-Rniz et al. (Plenum, 
New York, 1992).
%
\bibitem{30} 
D. E. Wolf, p.215 in {\it Scale Invariance, Interfaces, and 
Non-Equilibrium Dynamics}, edited by A. McKane et al. 
(Plenum, New York, 1995).
%
\bibitem{31}
P. Punyindu and S. Das Sarma, Phys. Rev. E {\bf 57}, R4863 (1998).
%
\bibitem{32}
H. Kallabis, L. Brendel, J. Krug and D. E. Wolf, Int. J. Mod. Phys.
B {\bf 11}, 3621 (1997).
%
\bibitem{33}
M. Rost and J. Krug, J. Phys. I {\bf 7}, 1627 (1997).
%
\bibitem{34}
S. Das Sarma, S. V. Ghaisas and J. M. Kim, Phys. Rev. E {\bf 49},
122 (1994).
%
\bibitem{35}
M. Schroeder, M. Siegert, D. E. Wolf, J. D. Shore and M. Plischke,
Europhys. Lett. {\bf 24}, 563 (1993).
%
\bibitem{36}
C. Dasgupta, S. Das Sarma and J. M. Kim, Phys. Rev. E {\bf 54},
R4552 (1996);
C. Dasgupta, J. M. Kim, M. Dutta and S. Das Sarma, Phys. Rev E
{\bf 55}, 2235 (1997).
%
\bibitem{37}
S. Das Sarma and P. Punyindu, Phys. Rev. E {\bf 55}, 5361 (1997);
S. Das Sarma, P. Punyindu and Z. Toroczkai, unpublished.
%
\bibitem{38}
S. F. Edwards and D. R. Wilkinson, Proc. Roy. Soc. London A
{\bf 381}, 17 (1982).
%
\bibitem{39}
F. Family, J. Phys. A {\bf 19}, L441 (1986).
%
\bibitem{40}
J. Krug, M. Plischke and M. Siegert, Phys. Rev. Lett. {\bf 70}, 
327 (1993).
%
\bibitem{41}
S. Das Sarma, P. Punyindu and Z. Toroczkai, Phys. Rev. E (in press).
%
\bibitem{42}
P. Punyindu, Z. Toroczkai and S. Das Sarma, Phys. Rev. B 
{\bf 64}, 205407 (2001).
%
\bibitem{43}
Y. Lee, I. Kim and J. M. Kim, J. Kor. Phys. Soc. {\bf 33}, 190 (1998).
%
\end{thebibliography}
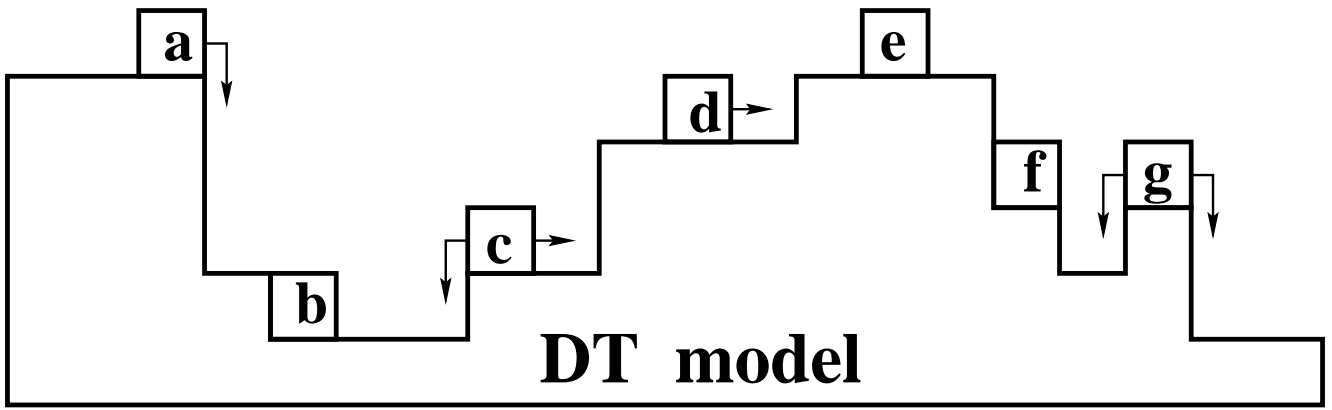
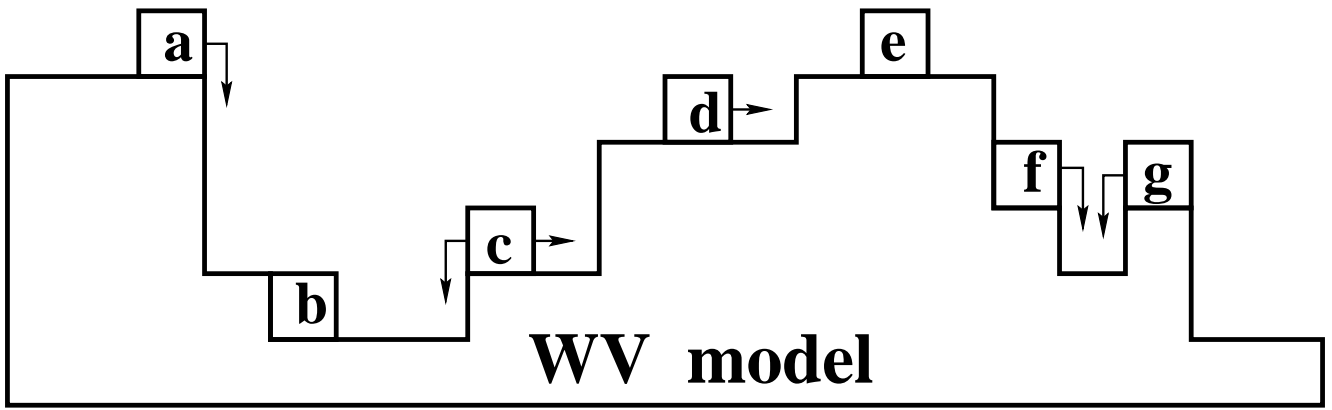
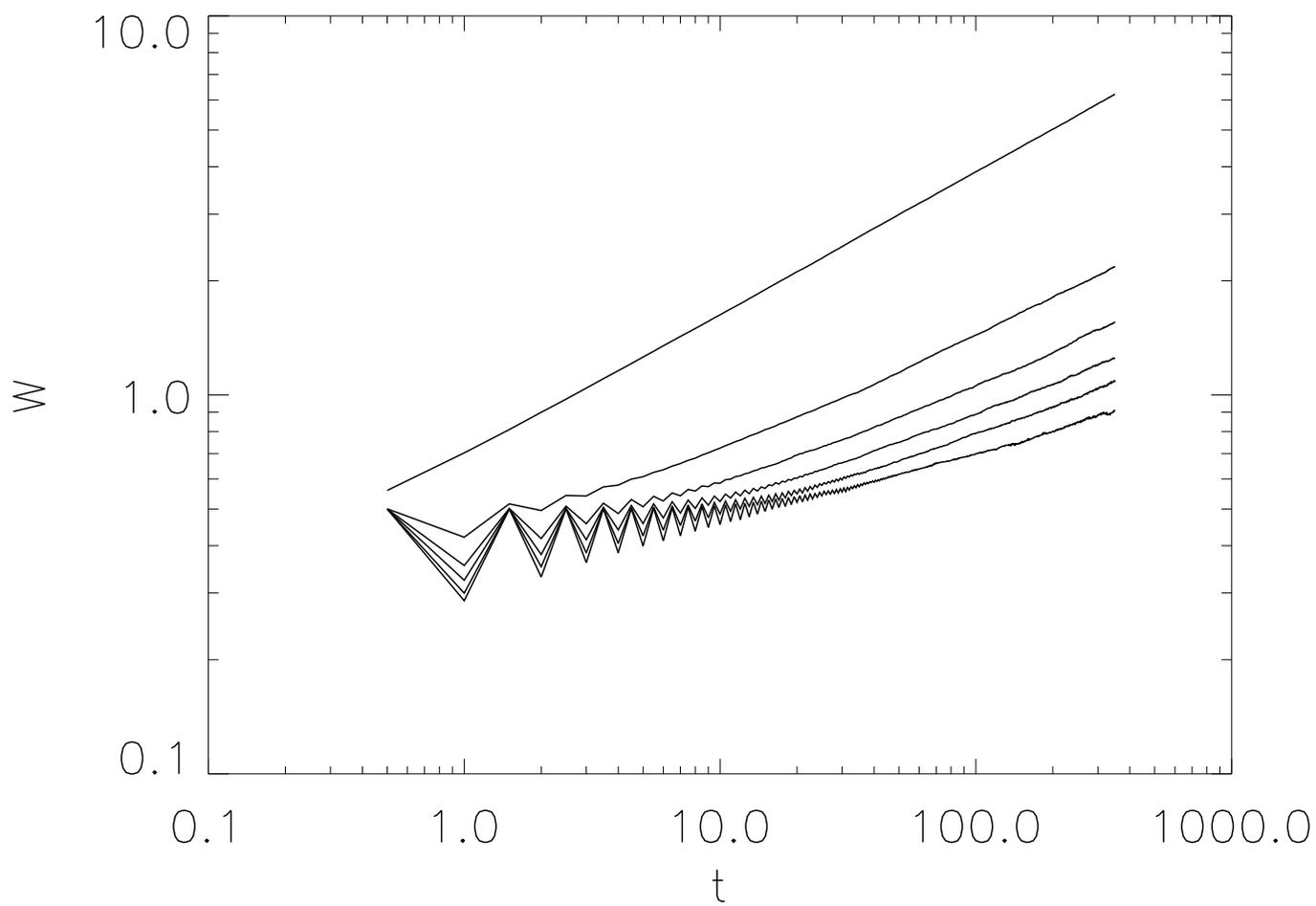
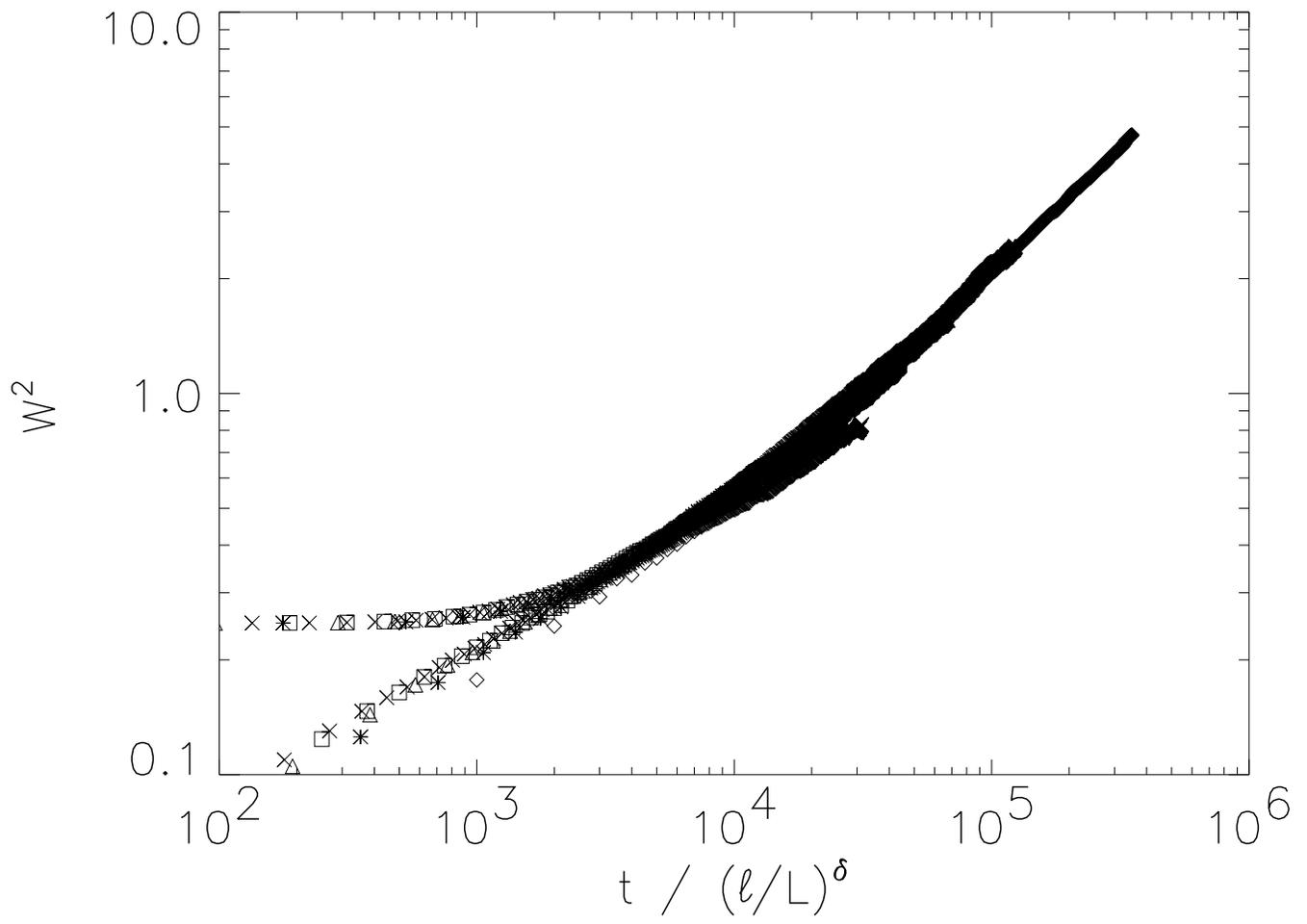
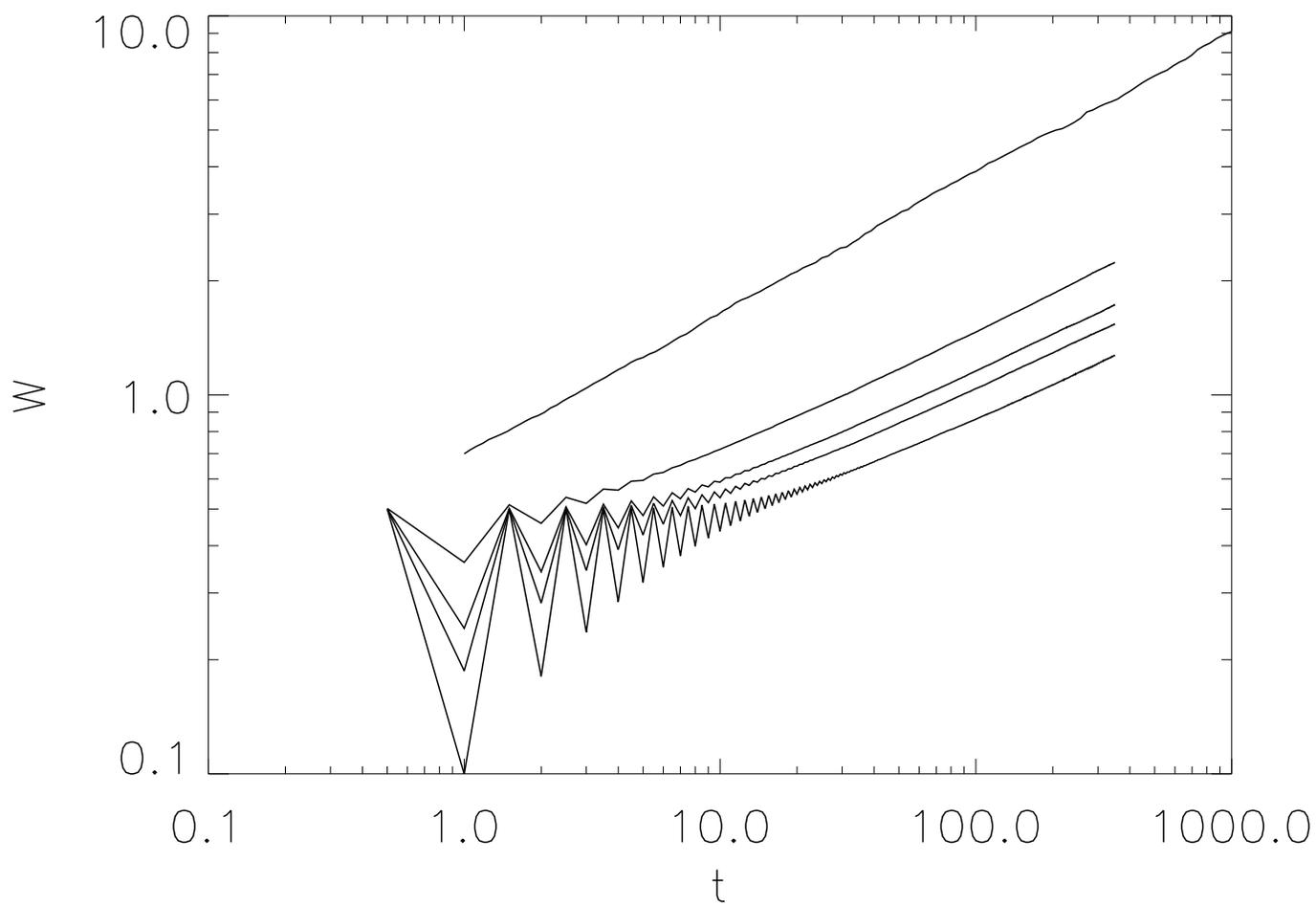
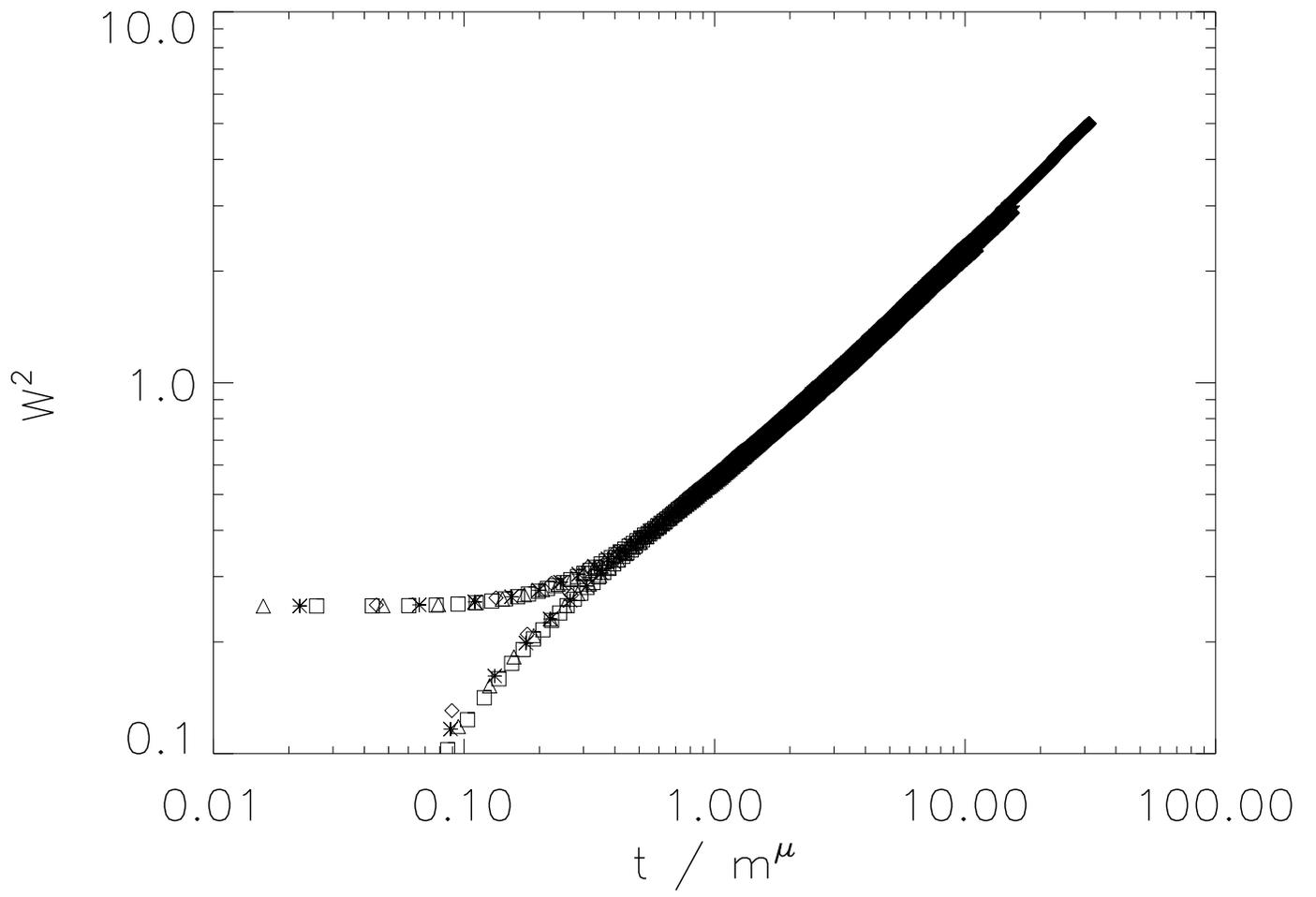
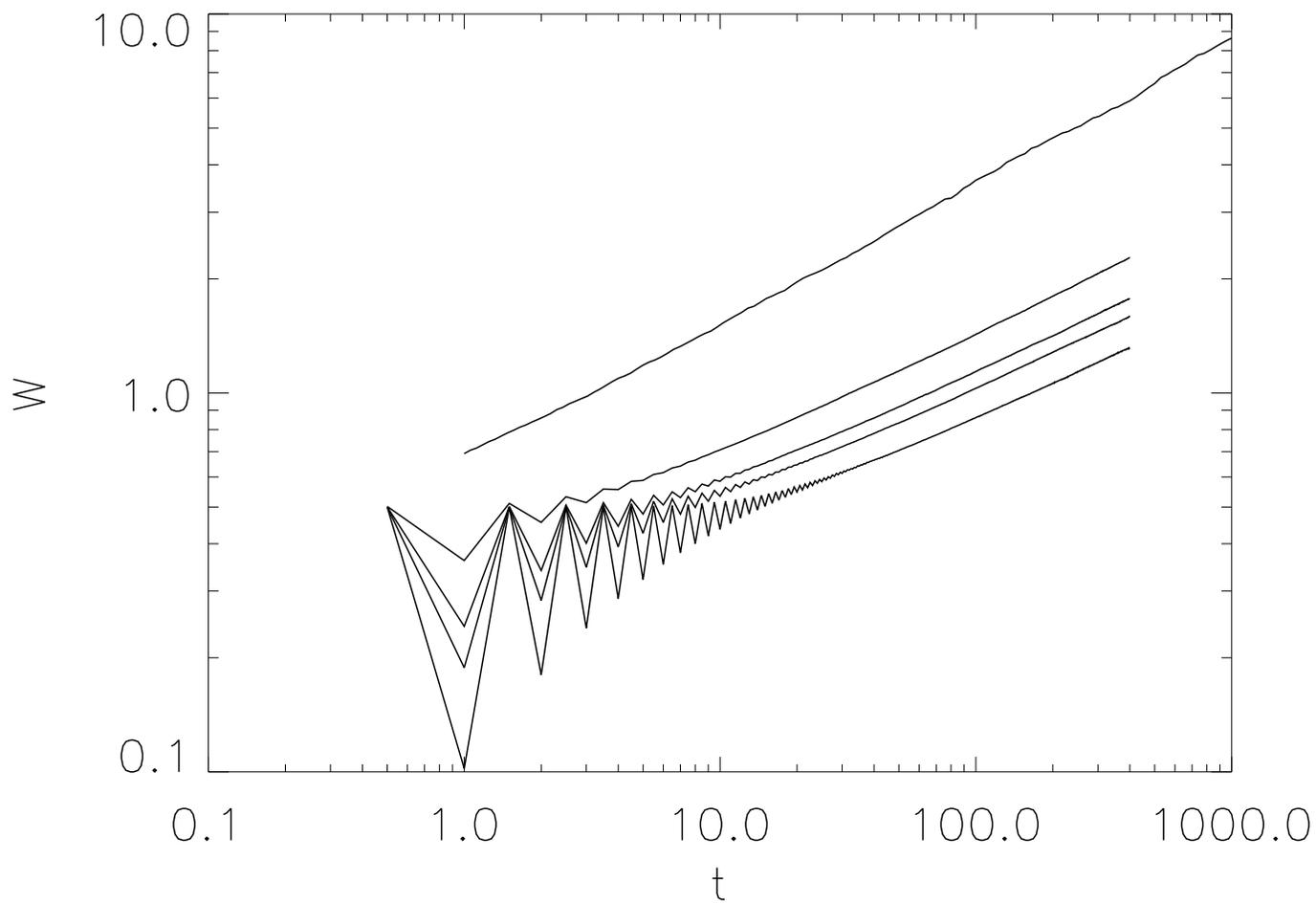
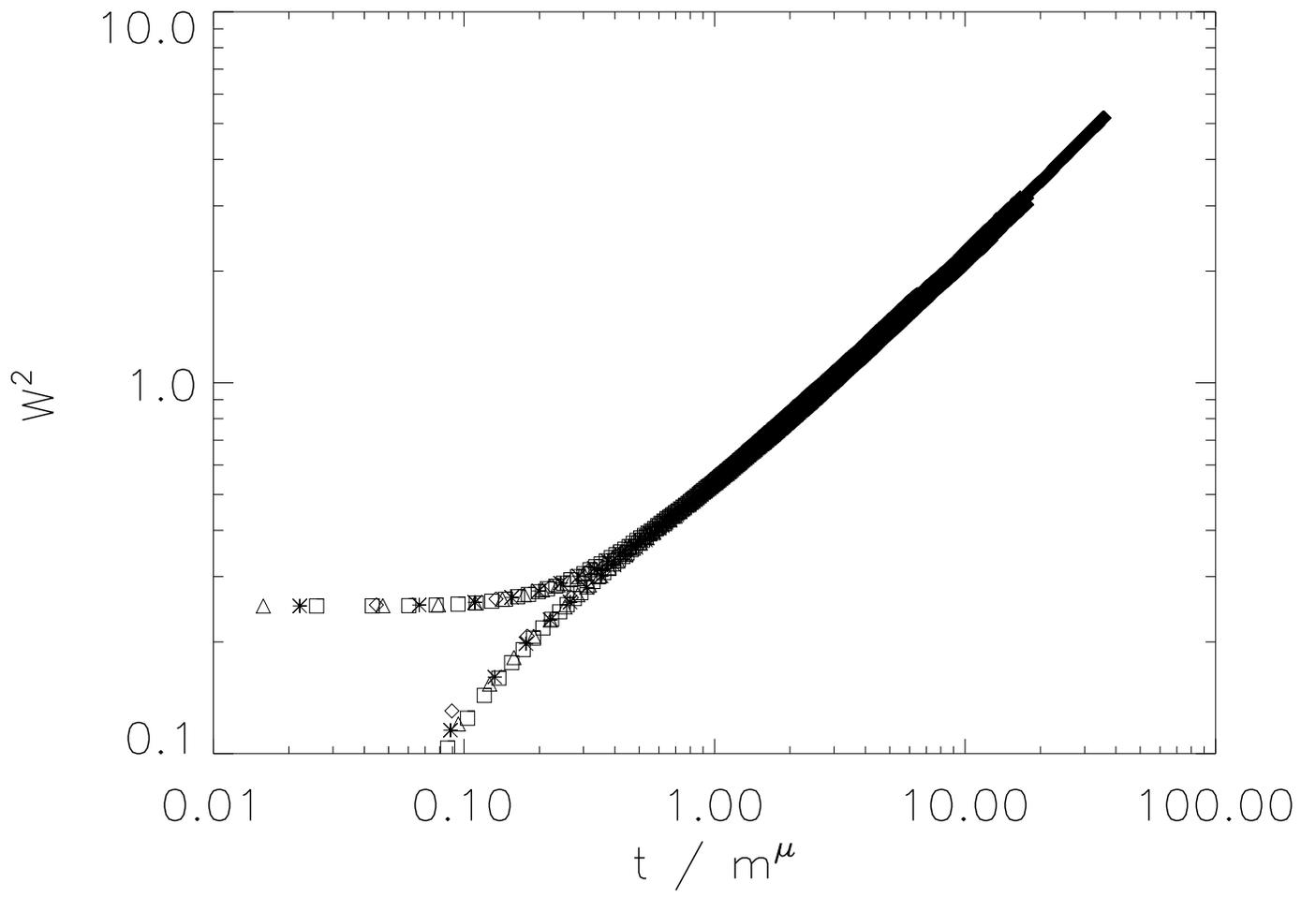
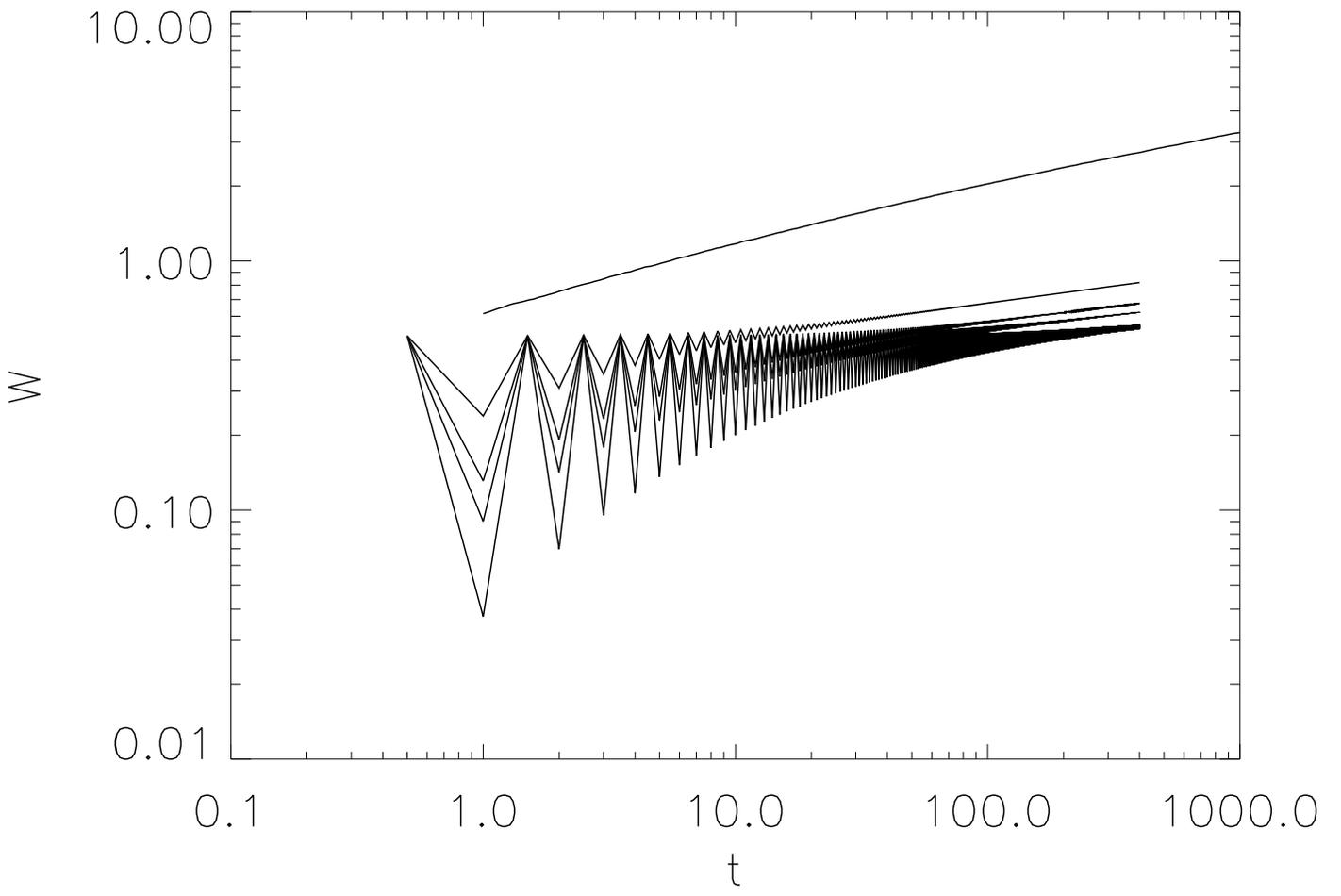
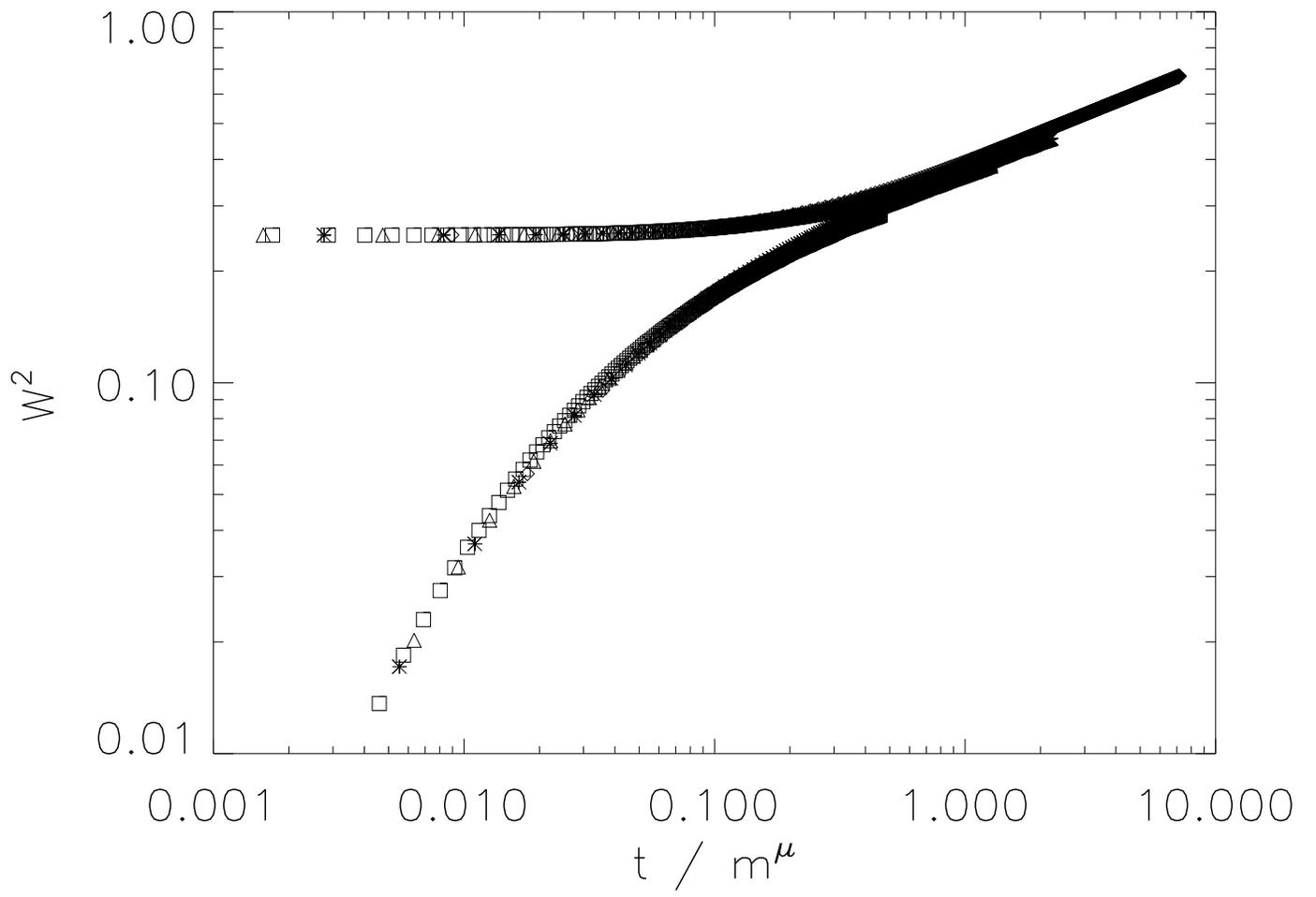
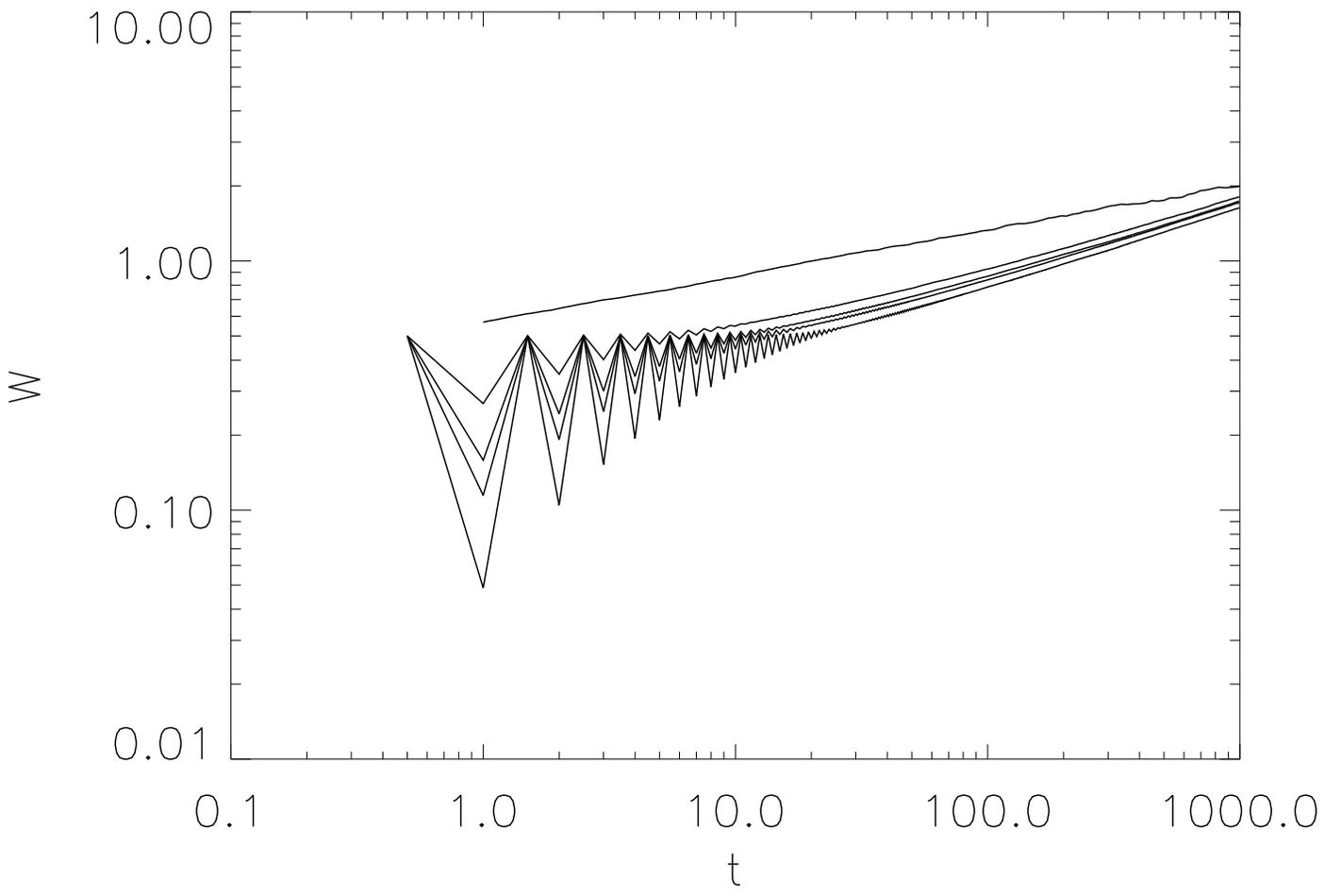
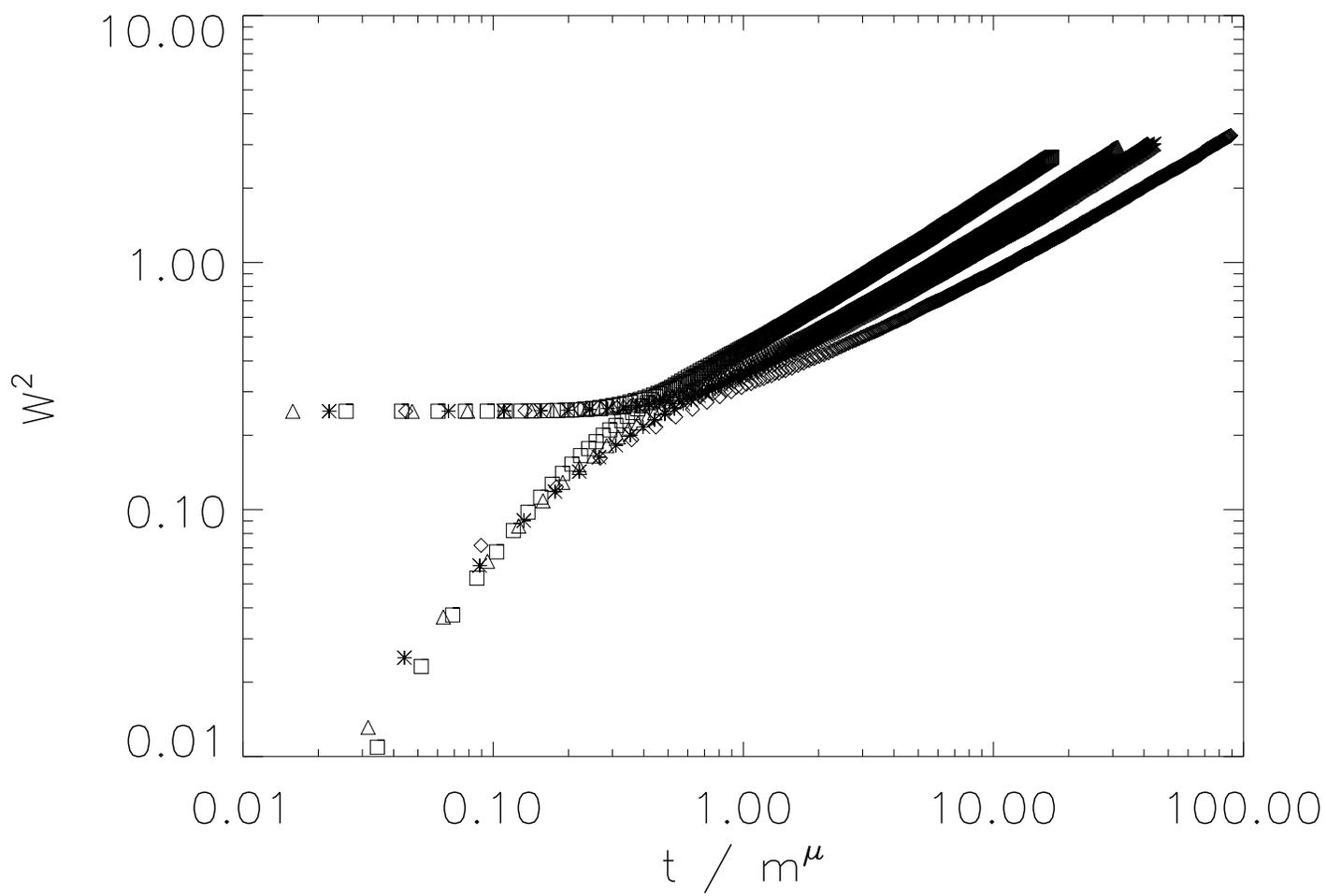
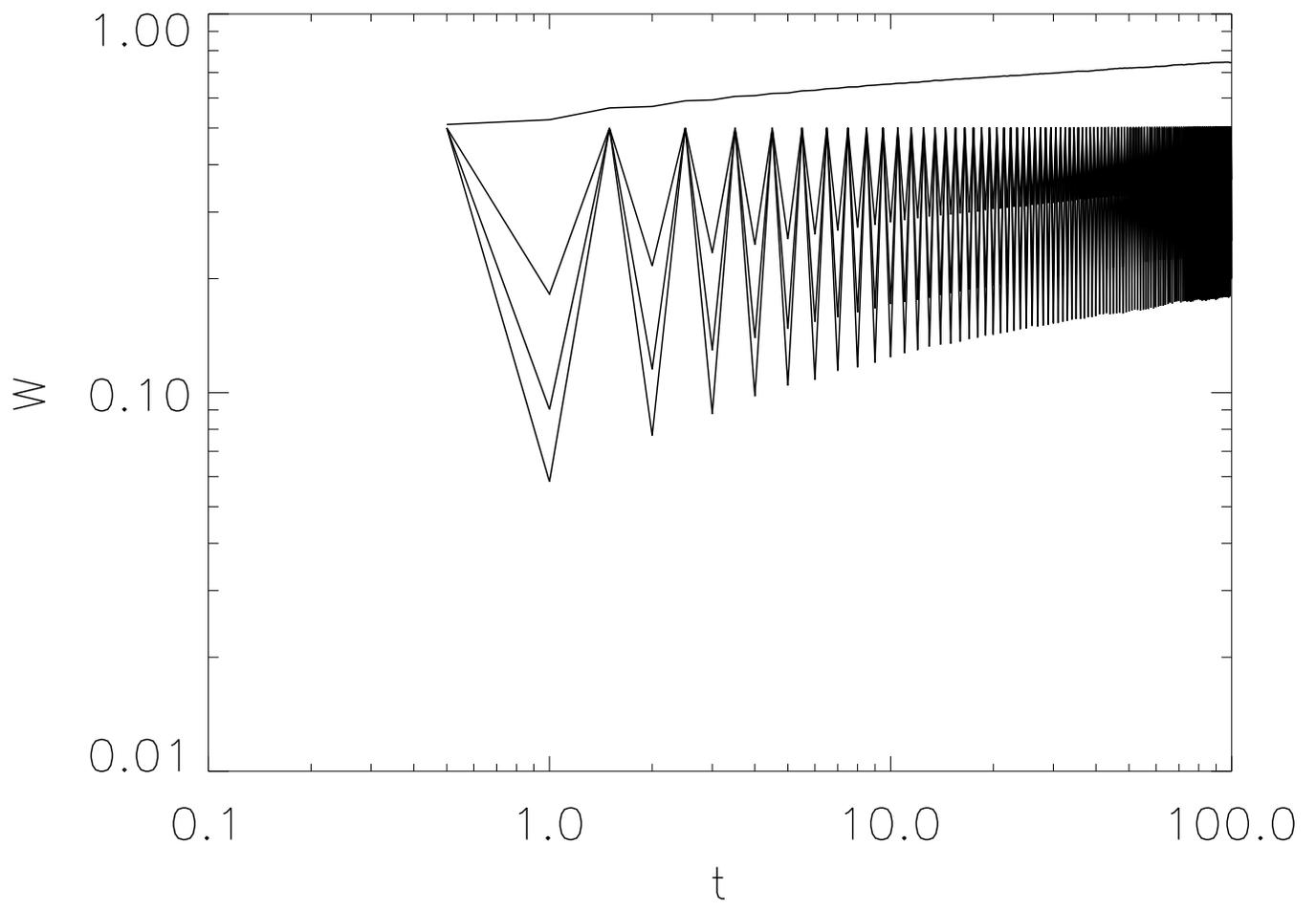
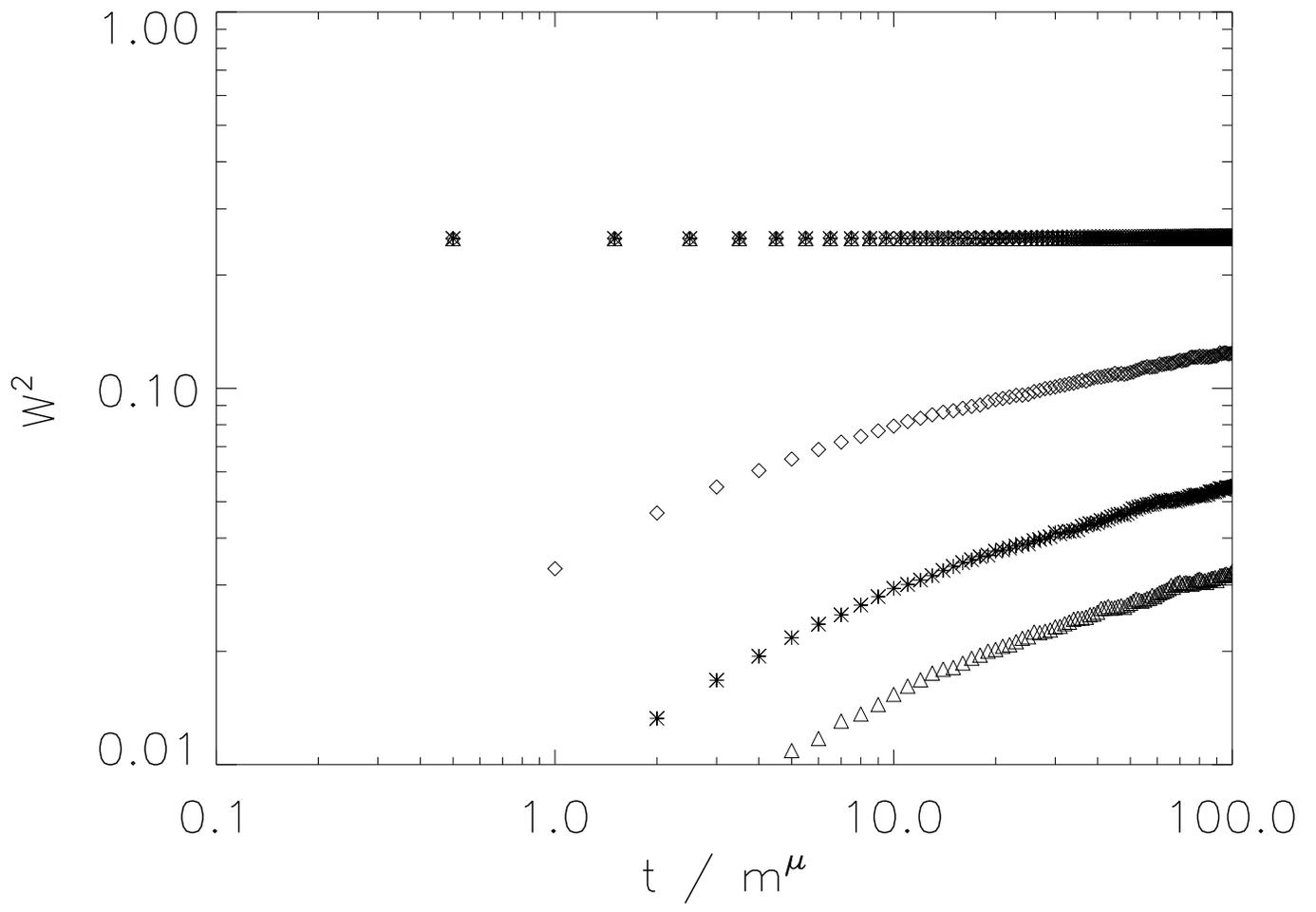
\end{document}